\documentclass[aps,onecolumn,groupedaddress,floatfix]{revtex4}
\usepackage[colorlinks=true,citecolor=red,filecolor=green,linkcolor=blue,pdfnewwindow=true]{hyperref}
\usepackage{amsmath} 
\usepackage[version=4]{mhchem}
\usepackage{color}
\usepackage{graphicx}
\usepackage{makeidx}
\usepackage{bm}
\usepackage[title]{appendix} 
\usepackage{tabu}
\usepackage{float}
\usepackage{caption}
\usepackage{subcaption}
\usepackage{hyperref}
\usepackage[normalem]{ulem}
\usepackage{natbib}
\usepackage{threeparttable}
\usepackage[dvipsnames]{xcolor}

\hypersetup{urlcolor=blue}

\makeatletter
 
\newcommand{\Rmnum}[1]{\expandafter\@slowromancap\romannumeral #1@}
\makeatother
\begin{document}
\title{Quantifying biofilm-virulence index to predict antifungal resistance in \textit{Candida albicans}}
\author{Nikhil Ujlayan$^{1,2,a}$}
\author{Teena Singh$^{1,2,a}$}
\author{Vanshika Dhama$^{1,2,a}$}
\author{Harsh Pratap Singh$^3$}
\author{Mahesh Kumar$^{2,*}$}
\author{R K Brojen Singh$^{1}$}\email{brojen@jnu.ac.in}
\thanks{$^a$ Equally contributing authors.}
\affiliation{$^1$School of Computational and Integrative Sciences, Jawaharlal Nehru University, New Delhi, India}
\affiliation{$^2$Department of Microbiology, JV College, CCS University, Baraut (Baghpat)-250611, India.}
\affiliation{$^3$Instruments Research Development Establishment, Dehradun-248 008.}

\begin{abstract}
\noindent\textit{Candida albicans} is a commensal microorganism that causes opportunistic infections, such as oral candidiasis, vaginitis affecting females, newborns, and immunocompromised patients. Biofilm formation can lead to a commensal organism becoming a life-threatening organism by introducing antifungal resistance. The experiment we did combines crystal violet staining for biofilm biomass and CFU counts to statistically construct an additive BVI model by analysing the experimental data. Our study on the data proposes a Biofilm-Virulence Index (BVI) as a novel and quantitative parameter for assessing antifungal drug resistance in \textit{Candida albicans}. The effect of the drugs on inhibition zone diameter is twofold, first, linear  increase with time during early biofilm formation, second, stabilizing in later phases and correlating directly with virulence.  Most BVI values remained in the mild infection range, indicating successful virulence reduction by antifungal drugs. The BVI model model combines the study of biofilm and viable cell count in a single parameter. So, this makes comparison between samples easier during biofilm analysis. Findings suggest that combination of CFU and biofilm measurement may improve interpretation of antifungal response in \textit{Candida albicans}. This approach could be useful in future experimental studies investigating biofilm associated resistance.\\

\noindent\textbf{Keywords:} \textit{Candida albicans}; BVI; CFU; Antifungal drugs; Statistical analysis. 
\end{abstract}

\maketitle

\section{Introduction}
\noindent\textit{Candida albicans}, a well-known species of fungi, is an opportunistic fungal pathogen that is commonly found as a part of the normal microbiota on human mucosal surfaces, including the mouth, gastrointestinal tract, and genital regions.\cite{Lamoth} \textit{C. albicans} becomes an opportunistic pathogen under uncommon conditions,\cite{pappas} reduced immune competence or imbalance of the competing bacterial microflora.\cite{Bougnoux me} Major fungal infections lead to the death of 1.6 million people every year. A systematic review based on approximately 434 published papers suggested that around 60 million Indians 4.1 are affected by serious fungal infections.\cite{beigi} It causes mucosal infections like oral candidiasis or vaginitis. Candidiasis can also occur in healthy individuals, but mainly affects newborns, Immunocompromised individuals that are having diabetes, pregnancy or newborns are readily affected or get it \cite{Chow BDW}, \cite{Ali GY} and the people taking broad-spectrum antibiotics or undergoing chemotherapy due to changes in their resident microbiota, facilitating conditions that contribute to the excessive increase of \textit{C. albicans} in their body.\cite{Lamoth F} Changes in the epidemiological landscape of invasive candidiasis. \textit{Candida albicans} leads to 70\% of fungal infections that lead to life-threatening invasive infections in recent decades. Despite proper treatment, the mortality rate is up to 40\% mostly in hospitalised conditions.\\

{\noindent}Candidiasis is a prevalent fungal infection, primarily caused by yeasts from the Candida species, with \textit{Candida albicans} being the most common culprit. It affects multiple body sites, including the mouth, vagina, skin, and oesophagus,\cite{Taff} and, in severe cases, the bloodstream, where it becomes invasive candidiasis, a serious threat to critically ill and immunocompromised individuals.\cite{Ray} Millions of women globally experience vulvovaginal candidiasis, while oral thrush is persistent in infants, those with compromised immune systems, denture wearers, people with uncontrolled diabetes, and substance users. Although most Candida species inhabit the human microbiome harmlessly,\cite{Reff} the risk of infection rises sharply with shifts in acidity, hormonal changes, the use of antibiotics or steroids, uncontrolled diabetes, and weakened immunity, often resulting in overgrowth.\\

{\noindent}Clinical symptoms of candidiasis can be severe, including intense itching and soreness, accompanied by thick white discharge in vaginal infections and painful, creamy white lesions in oral thrush, which can sometimes make eating or
swallowing difficult.\cite{De rose}, \cite{Fattani} Most cases are treatable with topical or oral antifungal medications; \cite{Tournu}however, emerging drug resistance, poses significant management challenges and threatens global health, compounded by limited diagnostic accessibility in many regions. Preventive strategies focus on proper hygiene, effective oral care, diabetes management, cautious antibiotic use, and appropriate use of steroid inhalers. At the same time, WHO guidelines emphasise avoiding synthetic fabrics, changing clothing after exercise, and refraining from douching or harsh vaginal products. Recognising the growing burden of antifungal resistance and the global impact of candidiasis, the World Health Organisation launched its first fungal priority pathogens list in 2022 ,\cite{Miceli}, \cite{Kabir}, \cite{Krause}, \cite{Brown} and
published targeted recommendations for diagnosis and treatment of Candida infections in 2024. These measures represent a global response aimed at enhancing awareness, advancing research, and improving health system preparedness to combat the challenges posed by Candida species, particularly in vulnerable populations. Negative Effects: \textit{Candida albicans} is responsible for a wide range of infections, from superficial mucosal infections like oral thrush and vulvo vaginal candidiasis to life- threatening invasive candidiasis, particularly in immunocompromised and critically ill patients. These infections cause substantial morbidity and mortality, with invasive candidiasis mortality rates reaching up to 50\% in hospital settings.\cite{Denning} The disease burden includes increased healthcare costs and resource utilisation worldwide. Emerging antifungal resistance, particularly in certain populations and geographic regions, complicates treatment and worsens health outcomes. Delays in diagnosis and the absence of an effective vaccine further amplify the negative impact on population health. Positive Effects: Conversely, \textit{Candida albicans} commonly exists harmlessly as part of the human microbiome on skin and mucosal surfaces. As a commensal organism, it plays a critical role in training and modulating the immune system, especially mucosal immunity, which helps protect the host against both fungal and bacterial pathogens. Colonisation by \textit{C. albicans} can contribute to immune homeostasis and microbiome balance. Ongoing research into its virulence mechanisms informs the development of better antifungal therapies and public health strategies. Most isolates remain susceptible to standard antifungals, enabling effective treatment in many cases.\\

{\noindent}\textit{C.albicans} infections are medically distinguished into two subunits: Mucosal infection and another one is Systemic. The fungal cells cause infection in the host, making its cells drug-resistant, and then evading the host immune system, especially during systemic candidasis. And due to an increasing range of yeasts that are showing resistance to antifungal drugs are an important factor in contributing to candida species infections in humans having long-term infection treatment in hospitals, such as HIV and Cancer patients\cite{Pfaller}.\cite{Tan} The biofilm is an important reason for the mediation of fungal drug resistance. Biofilms are structured microbial communities embedded within an extracellular matrix, whereby the embedded cells can show adhesion to each other. These cells within the biofilm produce the extracellular polymeric substances (EPS), which are a combination of proteins, extracellular polysaccharides, lipids and DNA.\cite{Tsui} Extracellular matrix protects the fungi from immune reactions and antifungal agents. \textit{Candida albicans} has a large tendency for formation of biofilms on abiotic and host surfaces by adhesion phenomena to catheters and medical devices \cite{Tumbarello} This shows its high virulence, and this biofilm(surface-associated communities of cells) formation leads to resistance to the antifungal therapy by the immune system and environmental factors.Several studies of \textit{Candida albicans} focused on Candida biofilm biomass, metabolic activity or viable cell count independently, Although no one study integrative mathematical work combining these parameters into quantitative virulence index
Most existing studies evaluate virulence-associated factors separately, limiting standardized comparison of infection severity and antifungal response across experimental conditions.\\

{\noindent}In this work, we provide and study a standardaized quantitative model which integrate both biofilm and CFU count for \textit{Candida albicans}. Further, we proposed a parameter \textit{biofilm virulence index} (BVI), which could be used as an index for quantitative evaluation of biofilm with virulence and antifungal response. Therefore, in this present study, we aim to develop and evaluate an integrative additive Biofilm Virulence Index (BVI) model using biofilm biomass and viable cell count measurements for quantitative assessment of \textit{Candida albicans} virulence and antifungal susceptibility. In the section II, the methods regarding experimental framework and statistical analysis are described. Experimental results and analysis of the experimental data are described in section III. We, then, summarised the results in the conclusion.

\section{Methods}

\subsection{Sabouraud Dextrose Agar(SDA) Method}
{\noindent}This study is engaged on the laboratory based experiment to examine and analysis the growth and antifungal susceptiblity of \textit{Candida albicans}. The isolation of the \textit{C.albicans} was cultured on Sabouraud Dextrose agar(SDA). It is a agar medium which is used for the cultivation of fungi and yeast. This media which is specially used for fungal growth  was named  by the french physician Raymond Sabouraud in 1892. The composition  used for making SDA contains dextrose (sugar), peptone (source of nitrogen and vitamins), and agar (which is for solidifying the media). The suitable pH for the media is used around (5.4-5.6) which encourage the growth of \textit{C.albicans} for the selective study growth and incubate under room temperature at 37$^{\circ}$C for 24-48 hours on SDA.\cite{Soll}, \cite{Thomas}  Antifungal activity was done by using disc diffusion method in which the antifungal agents were Itraconazole, Griseofulvin were used for the antifungal activity. We explain here the experimental procedure using the above mentioned techniques on the fungal microbial species
\textit{Candida albicans}. The pictorial procedure is given in the Fig. 1 (a).

\noindent\textbf{Media Preparation:}
Sabouraud Dextrose Agar (SDA) contains 40 g of dextrose per litre 10 g of peptone, and 15 g of agar. This makes the favourable condition for growth of \textit{Candida albicans}. To prepare it, start by adding 40 g of dextrose which also called glucose, 10 g of peptone and 15 g of agar into 1 litre of distilled water. Now boil everything until it’s fully dissolves. After that sterilise the media in an autoclave at 121°C for 15 mins, now let it cool down to 45 to 50℃. Before pouring add a pinch of streptomycin and streptopenicllin in ratio of (1:1) to suppress the growth of Gram + and Gram - bacteria and to get a pure culture of \textit{Candida albicans}, and pour it into the sterilised petri dishes to make a layer of solidify.

\noindent\textbf{Inoculation and culture:}
After preparing SDA petri plates, \textit{Candida albicans} is inoculated by picking up a few colonies from an overnight broth using a sterile loop. After that streak the inoculum across the agar surface in a zig zag pattern and spread it evenly by using the spreader for isolated colonies. After streaking method seal the petri plates with paraffin tape which creates a air tight seal on the petri plates to get contaminated. Now incubate the plates upside down at 35-37℃ for 24 to 48 hrs. Now allowing creamy white and smooth colonies to form.

\noindent\textbf{Inoculum preparation:}
To prepare the inoculum for \textit{Candida albicans} pick 4-5 separate creamy white colonies from a 24 hrs SDA culture with a sterile loop and collect in microcentrifuge tubes and mix them into sterile saline until it looks milky for mixing well use vortex shaker to mix it well properly.

\noindent\textbf{Antifungal test setup:}
For the antifungal test setup on \textit{Candida albicans} use a 75 well plates and add 10µL medium to each well. Then make two folds dilutions of the antifungal drug, such as Itraconazole and Griseofulvin across the wells from high dose like 64µg/ml down to low one like 0.125 µg/ml. Now setup controls drug only wells for sterility medium plus cells for growth check and plain medium as a blank. After drug dilution, the plates were sealed and incubated at 35°C for 24 h under static conditions.

\noindent\textbf{Incubation and reading:}
Incubate at 35℃ for 24hrs it can be extended to 48 hrs it depends on the aerobic conditions. Read the Minimum
Inhibitory Concentration (MIC) is the lowest concentration with greater than 50

\subsection{Crystal Violet Staining Method}
{\noindent}The procedure for the crystal violet staining in \textit{Candida albicans} is hereby in some diﬀerent steps, which we used to stain the \textit{C. albicans}. The materials that are used in preparation first are \textit{Candida albicans}, a freshly cultured 1\% crystal violet
solution to see the cells under the microscope, sterile PBS (phosphate buﬀer saline) with 95\% ethanol, microtiter plates or microwell plates are required and a plate reader. Then, the procedure for biofilm formation is to be done as follows steps Firstly, with the help of an inoculating loop, pick the sample from the freshly grown \textit{C. albicans} and inoculate it in the wells. Take a very low amount of the sample, $1 \times 10^{7} \ \text{cells}/\text{ml}$ Then incubate the microwell plate in the incubator for the next 24 hours at a temperature of 30-37°C to form the biofilm growth. The second step is used to wash the culture media with the help of saline water(PBS) to remove the non-anchorage-dependent cells. This process is done twice, and the cells that require a solid surface as a tissue culture vessel in order used to grow and proliferate. After washing the culture media, the drying process is followed, allowing the wells to dry for 45 minutes to fix in the wells for the staining process. After the drying step, the staining process is started first 1\% crystal violet solution, 2-3 drops in each well. After adding the crystal violet solution, incubate the microwells for 20-45 minutes at room temperature. After the incubating period, wash the microwells with saline water 3-4 times to remove the excessive stain in the next step, then add 4-5 drops of ethanol 95\% to each well and leave for the next 45 minutes to solubilise the bound dye. There are some preconditions to execute the procedure, including the use of triplicates and negative controls for reliable quantification. Microwells should be free of yeast. Handle the microwells gently throughout the washing steps to avoid dislodging biofilm.\cite{Gulati}, \cite{Chandra}.

\subsection{Statistical Analysis}
{\noindent}We used linear regression analysis to study the dynamics of the inhibition zone diameter dynamics of the Candida albicans for two different antifungal drugs and their growth behaviour. We used XmGrace plotting software and Microsoft Excel for the data representation, analysis and the goodness of the fit was understood from the $R^2-$values. Correlation and regression analyses was done using Pearson’s correlation tests using the plotting software.

\section{Results}

{\noindent}The experimental data using the experimental procedure on the growth of \textit{Candida albicans} described in the method section is analysed specially focussing on the dynamics of the growth and inhibition of this micro-organism affected by the two antifungal drugs Itraconazole and Griseofulvin. Another aim of this study is to propose a proper equation for calculating BVI derived from the experimental observations of the growth of this micro-organism. \\
\begin{figure}[h]
    \centering
    \includegraphics[width=0.9\textwidth]{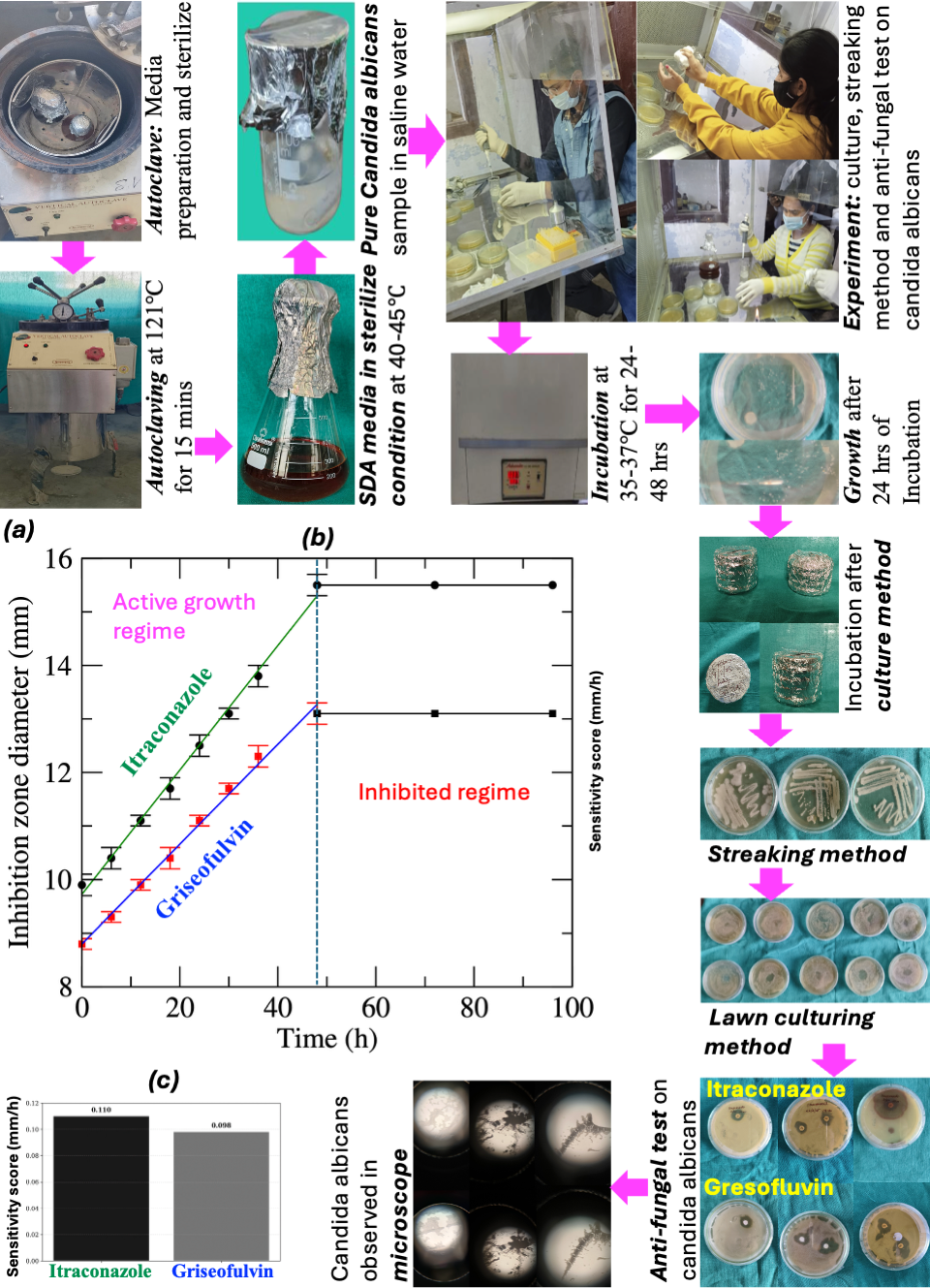}
    \caption{\textbf{Impact of the antifungal drugs on the growth of Candida albicans:} (a) Experimental set up and process of measuring inhibition zone diameters of the Candida albicans at different times within [0-100]hours for two antifungal drugs, namely, Itraconazole and Griseofulvin. (b) The plots indicate the dynamics of the growth of the Candida albicans in the presence of the antifungal drugs separately. Each filled circle is the mean of the measured inhibition zone diameters in the three samples with corresponding error bar with standard deviation. The straight lines on the data (green and blue colours) are the linear regression fits on the data points of the two drugs respectively. (c) The sensitivity plots of the two antifungal drugs on Candida albicans growth.}
    \label{fig1}
\end{figure}

\subsection{Antifungal drug susceptibility: Inhibition zone regression and sensitivity analysis}

{\noindent}We study the dynamics of the growth of the micro-organism \textit{Candida albicans} in the presence of two antifungal drugs, namely, Itraconazole and Griseofulvin separately to study the effectiveness of the drugs on this micro-organism. We took three different samples of the same experiment for each antifungal drug at a particular time (in hours) to calculate inhibition zone diameter (see the supplementary table). Then we took the mean of the three samples at each time of measurement with the corresponding standard deviation or error. We measured the inhibition zone diameter at different times within [0-100]hours for the two antifungal drugs separately. The results are shown in the Fig. 1 (b). Now, we use inhibition zone regression method, which is a quantitative statistical regression method on the experimentally obtained data, used to analyse the data to evaluate the effectiveness of antifungal agents by measuring and modelling the inhibition zone diameter of the clear area (zone of inhibition) around a drug-impregnated disk or well over time. \\

{\noindent}We, then, analyse the dynamics of the inhibition zone data using statistical analysis as shown in the Fig. 1 (b). The data clearly show two distinct regimes for both antifungal drugs, one active growth regime and the other maximal inhibited regime which is the saturated regime or plateau (inhibition zone diameter does not change with respect to time). In the active growth regime, we fitted the data with linear equation,
\begin{eqnarray} 
\label{fit}
x(T)=aT+b
\end{eqnarray} 
where, $x$ is the inhibition zone diameter with respect to time $T$ in hours. $a$ and $b$ are constants) separately to the two sets of the data points of the two drugs respectively. The goodness of the fits are found to be very good, $R^2=0.998$ for the \textit{Itraconazole} drug data set and $R^2=0.998$ for the Itraconazole and Griseofulvin data set respectively. This results indicate that the \textit{Candida albicans} grow linearly in the active growth regime with the slopes $a_i=0.116124$ ($b_i=9.72431$) and $a_g=0.0931495$ ($b_g=8.799$) when the drugs Itraconazole and Griseofulvin are infused respectively. If we represent $x_i$ and $x_g$ represent inhibition zone diameters of \textit{Candida albicans} affected by the drugs Itraconazole and Griseofulvin respectively, then the results indicate that $x_i>x_g$ for every time $T$ revealing the impact of Itraconazole is much more than that of Griseofulvin. Further, since $\frac{a_i}{a_g}\sim\frac{5}{4}$ the growth of the \textit{Candida albicans} is inhibited $5/4$ times faster than when Itraconazole drug is administered than when the drug Griseofulvin is administered. Even the maximal inhibited diameter of Itraconazole is much higher than that of Griseofulvin indicating inhibitory impact of Itraconazole is quite larger than that of Griseofulvin ($\frac{x_i^s}{x_g^s}\sim 1.18$).\\

{\noindent}Generally, in the context of inhibition zone regression analysis, the sensitivity is defined as the rate at which the diameter of the inhibition zone expands per hour. It tells how fast an antifungal drug inhibits the growth of \textit{Candida albicans}. From the sensitivity analysis of the data of the \textit{Candida albicans} growth (Fig. 1 (c)), the results show that Itraconazole has a higher sensitivity score (0.116 mm/hr) as compared to that of Griseofulvin (0.093 mm/hr). Hence, Itraconazole drug inhibits \textit{Candida albicans} microbial growth faster and is more effective than Griseofulvin drug.\\

{\noindent}Further, Itraconazole demonstrated larger and more stable inhibition zones (9.5 mm plateau at 24 hours; sensitivity score 0.116 mm/hr) than Griseofulvin (7 mm; 0.093 mm/hr), confirming higher potency and faster anti-biofilm action. The effective concentration was reached within 24–36 hours and remained stable thereafter. In contrast, comparative research finds Fluconazole, Amphotericin B, and similar antifungals often yield delayed plateaus and smaller inhibition zones, especially in mature and resistant biofilms\cite{Dawoud}, \cite{FCIMB} found that non-albicans Candida isolates are less susceptible, with narrower inhibition zones. Itraconazole has a higher sensitivity score (0.116 mm/hr) as compared to Griseofulvin (0.093 mm/hr). Itraconazole inhibits microbial growth faster and is more effective than Griseofulvin. While traditional studies do not consistently use sensitivity scores, their biofilm response kinetics are generally slower, emphasizing the value and novelty of a time-resolved, quantitative approach.

\begin{figure}[h] 
    \centering
    \includegraphics[width=1.0\textwidth]{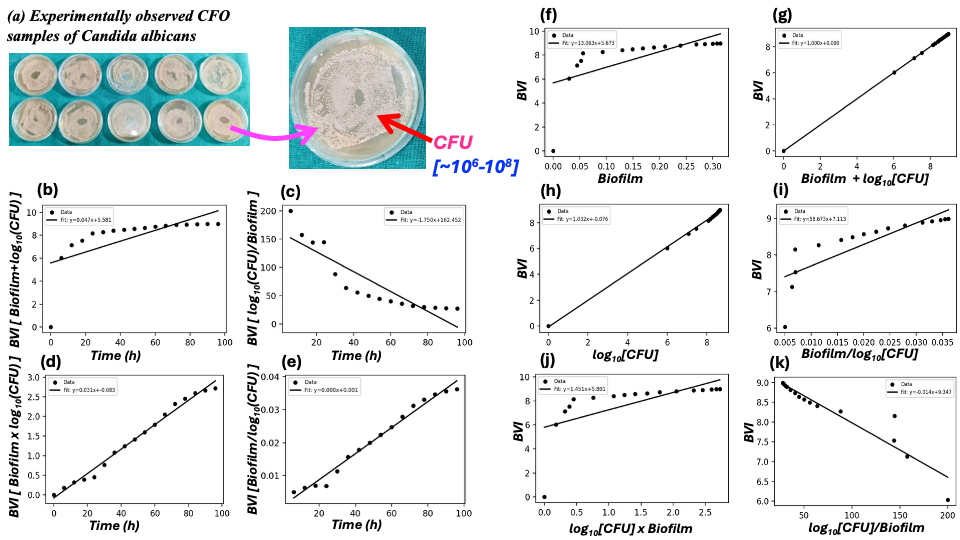}
    \caption{\textbf{Regression analysis of the experimental data on \textit{Candida albicans}:} (a) Experimental observation of \textit{Candida albicans} growth on the sample and then calculate CFU. (b)-(k) All possible plots between BVI and CFU as a function ot $T$. The filled circle points are experimentally calculated values as a function of $T$ in hours. The straight lines are the fits using linear regression analysis.}
    \label{fig2}
\end{figure}

\subsection{Modelling virulence factor expression and integrative equation}

{\noindent}In microbial study Biofilm Virulence Index (BVI) correlates to the biofilm formation which indicates a critical virulence factor \cite{Storms}. This BVI is generally defined as an indicator of infection severity, which depends on the measurable biofilm characteristics, specifically the biofilm biomass (\text{Biofilm Score}) and the number of viable cells within the biofilm (\(\log_{10}(\text{CFU})\)) \cite{Ouyang}. However, there is no exact general expression relating BVI with CFU satisfying for all micro-organism growth dynamics. In order to establish this model equation connecting BVI and CFU for the growth dynamics of \textit{Candida albicans}, we statistically analysed the experimental data extensively.\\

{\noindent}Now, we plotted all different possible plots which correlate BVI, CFU and time as shown in Fig. 2 (b)-(k). Then we did regression analysis of all the data to obtain the most accurate correlating equation between BVI and CFU. The plots and their regression analysis show that the results which connects BVI linearly with $log_{10}(CFU)$ (Fig. 2 (h), (g)) are found to be best correlated results out of all the fitted data (Fig. 2 (b)-(k)). The further analysis indicates that the best relating equation between BVI and CFU is found to be given in Fig. 2(g), and the corresponding equation is given by,
\begin{eqnarray}
\label{bvi}
BVI = a \times Biofilm ~Score + b \times \log_{10}(CFU) + c,
\end{eqnarray}
where, $a$, $b$ and $c$ are are the coefficients representing the relative weight or contribution of biofilm biomass and viable cell count to the overall virulence. Now, fitting with this equation \eqref{bvi} to the experimental data, we found that $a = 0.997\sim 1$, $b = 1.000$, and $c = 0.00004$ (comparatively very small). This regression analysis indicates an almost equal and linear contribution from both biomass and cell viability to biofilm virulence. The intercept is negligibly small (near zero), meaning the equation passes close to the origin and hence, the BVI increases primarily due to these two factors, which is given by the following additive linear equation, 
\begin{eqnarray}
\label{bvi1}
BVI \sim Biofilm ~Score + \log_{10}(CFU)
\end{eqnarray}
This additive linear model indicates that biofilm virulence is well predicted by combining biofilm mass and live fungal cell density, allowing clear quantification of infection severity over time. This model equation showed the strongest, linear, and biologically intuitive relationship throughout the entire biofilm lifecycle of the \textit{Candida albicans}. Alternative models, including multiplicative and ratio-based plots, introduced interpretive complexity or missed key features of biofilm dynamics, often producing complex or misleading patterns. In contrast, the additive equation effectively integrated both biofilm biomass and viable cell count, facilitating clear tracking of infection severity, robust comparison across treatments, and meaningful laboratory assessment. Hence, in general the model equation (3) provide the simplified additive model, and we propose this equation as a quantitative additive model for biofilm virulence assessment in \textit{Candida albicans} for biofilm virulence quantification in experimental \textit{Candida albicans} research to the experimental data, and we considered this equation as the standard model for biofilm virulence quantification in experimental \textit{Candida albicans} research. \\

{\noindent}The combination of biomass and viable cells in a single BVI index generally show that higher cell numbers and matrix production may lead to the increased in infection severity and resistance. However, most conventional studies \cite{Jacob,Shrief,Atiencia} reported that these virulence factors as separate measurements, lacking a unified index and do not give a direct index for analysis. Hence, our proposed integrative framework for calculating BVI represents a methodological advance, offering consistent risk stratification and direct cross-model comparison.

\subsection{Infection Severity Categorization and Biofilm Quantification}
{\noindent}The Biofilm Virulence Index (BVI) introduced in this study integrates crystal violet biofilm scoring and viable cell count (log10CFU), providing a quantitative assessment for \textit{Candida albicans} infection severity. Infection categories are defined as mild $\text{BVI} < 8$, moderate (8–11), and severe $> 11$. Most experimental results in this study fell into the mild category, with rapid early BVI increase and plateau, indicating effective control by antifungal agents. Previous work by \cite{Kuhn} compared Candida species and established that \textit{Candida albicans} forms larger, more drug-resistant biofilms, identifying high biomass as a marker for increased virulence and resistance.\cite{H.T} Validated biofilm measurement methods, highlighting metabolic activity and overall biomass as reliable indicators of pathogenic risk, though without a composite index like BVI. Other studies report severity and risk using separate measurements of biomass or cell count, indicating the value of integrated quantification for standardizing comparisons. This study’s BVI model thus advances standardization in biofilm quantification, providing direct, continuous assessment of infection risk an improvement over the traditional use of isolated parameters.

\section{Conclusion}
{\noindent}We studied antifungal drugs (Itraconazole and Griseofulvin) activities on the fungal microbe Candida albicans and proposed an integrative model equation to measure and stratify fungal virulence. The impact of the drugs on the fungus is found to be twofold, first the linear growth of the inhibition zone diameter of the fungus within a certain range of time, and second, to attain a maximum inhibited zone diameter. Hence, both the antifungal drugs inhibit the fungal microbial growth after a certain interval of time eventually.\\

{\noindent}Next, we proposed a model equation to measure BVI by systematic integrated quantitative analysis, which uses biofilm measurements via crystal violet staining and traditional CFU counts, giving the complete picture of \textit{Candida albicans} growth, antifungal susceptibility by drugs (Itraconazole and Griseofulvin), and infection risk. This statistical additive model is given by equation (2), which can explain the biofilm lifecycle trends with a strong, almost very accurate linear correlation \[R^2 > 0.99\], which provides clear justification with the proposed integrative additive model. Systematic regression analysis confirmed that Itraconazole demonstrates substantially more potent and rapid antifungal action, attaining a larger and sustained inhibition zone (plateau at 9.5 mm) and a larger sensitivity score (0.116 mm/hr) compared to Griseofulvin (plateau at 7 mm; sensitivity score 0.093mm/hr). Both agents reach their respective peak inhibitory effects within 36 hours and maintain stability through 96 hours. These findings show that Itraconazole in comparision to Griseofulvin, is the preferred antifungal drug for rapid and sustained inhibition of \textit{Candida albicans} biofilms. This further highlights that this parameter, BVI which is the integration of CFU, biofilm measurement, can be used as a quantitative marker for characterizing infection severity and therapeutic outcomes. Finally, various parameters come into play with a multifaceted, quantitative, biological approach that enhances the precision of antifungal susceptibility testing and provides a strong foundation for future research and clinical applications targeting Candida biofilm-associated infections.\\

{\noindent}Further, we propose that, the future studies should validate the proposed BVI framework across multiple Candida species and polymicrobial biofilms to evaluate broader clinical applicability. The proposed BVI model providing a simple and quantitative study of \textit{Candida albicans} by integrating biofilm biomass and viable cell count during assessment of antifungal susceptibility. The strong linear regression observed suggests that combination of quantitative analysis may raise and gives precise biofilm associated virulence analysis in \textit{Candida albicans}. Inhibition Zone regression with time also provides strong and additional insights into antifungal response kinetics as compared to conventional endpoint measure points. However, this study of \textit{Candida albicans} was done only under in vitro conditions and may require more validation using larger datasets and can include multi-dimensional parameters characterizing virulence of the organism. The proposed BVI model was developed using biofilm biomass and viable cell count measurements and therefore may not fully capture the complexity of host-pathogen interactions observed in clinical infections. Additional validation using larger datasets, clinical isolates, and broader virulence determinants is required to further establish the general applicability of the model. In future, studies may validate the proposed BVI model which include  multiple Candida species which are especially based on clinical datasets. Advance machine learning approaches will also led the improvement in the analysis of antifungal resistance and infection severity. Therefore, there can be additional virulence determinants such as extracellular matrix production, metabolic activity, hyphal transition, and biofilm-associated gene expression.\\

\noindent\textbf{Acknowledgements}\\
R.K.B.S. acknowledges DBT, BIC for financially support. \\

\noindent\textbf{Conflict of Interest}\\
The authors declare there is no conflict of interest.

\end{document}